\documentclass[prd,twocolumn,preprintnumbers,amsmath,amssymb,showpacs]{revtex4}

\usepackage{graphicx,bm}

\makeatletter
\def\graphicscale{\twocolumn@sw{0.3}{0.4}}
\def\graphicthreescale{\twocolumn@sw{0.3}{0.4}}

\begin{document}

\title{

Relevance of the axial anomaly at the
finite-temperature chiral transition in QCD

}

\author{Andrea Pelissetto}
\affiliation{Dipartimento di Fisica dell'Universit\`a di Roma ``La Sapienza"
        and INFN, Sezione di Roma I, I-00185 Roma, Italy}

\author{Ettore Vicari}
\affiliation{Dipartimento di Fisica dell'Universit\`a di Pisa
        and INFN, Sezione di Pisa, I-56127 Pisa, Italy}

\date{November 11, 2013}

\begin{abstract}

We investigate the nature of the finite-temperature chiral transition
in QCD with two light flavors, in the case of an effective suppression
of the U(1)$_A$ symmetry breaking induced by the axial anomaly,
which implies the symmetry breaking ${\rm U}(2)_L\otimes {\rm
  U}(2)_R\rightarrow {\rm U}(2)_V$, instead of ${\rm SU}(2)_L\otimes
{\rm SU}(2)_R\rightarrow {\rm SU}(2)_V$.  For this purpose, we perform
a high-order field-theoretical perturbative study of the
renormalization-group flow of the corresponding three-dimensional
multiparameter Landau-Ginzburg-Wilson $\Phi^4$ theory with the same
symmetry-breaking pattern.  We confirm the existence of a stable fixed
point, and determine its attraction domain in the space of the bare
quartic parameters.  Therefore, the chiral QCD transition might be
continuous also if the ${\rm U}(1)_A$ symmetry is effectively restored
at $T_c$.  However, the corresponding universality class differs from
the ${\rm O}(4)$ vector universality class which would describe a
continuous transition in the presence of a substantial ${\rm U}(1)_A$
symmetry breaking at $T_c$.  We estimate the critical exponents of the
${\rm U}(2)_L\otimes {\rm U}(2)_R\to {\rm U}(2)_V$ universality class
by computing and analyzing the corresponding perturbative expansions.
These results are important to discriminate among the different
scenarios for the scaling behavior of QCD with two light flavors close
to the chiral transition.

\end{abstract}



\maketitle


\section{Introduction}

At finite temperature ($T$) nuclear matter shows two different phases: a
low-$T$ hadronic phase, in which chiral symmetry is broken,
and a high-$T$ phase, in which chiral symmetry is restored and
quarks and gluons are unbounded
\cite{Wilczek-92,RW-93,Wilczek-00,Karsch-02,PW-84,GGP-94}.  Since the
$u$ and $d$ quarks are very light, a great amount of work has
been devoted to the study of QCD with $N_f$ light flavors, $N_f=2$
being the physically interesting case.  In this limit the QCD
Lagrangian is invariant under ${\rm U}(N_f)_L$ and ${\rm U}(N_f)_R$
transformations.  Since
\begin{equation}
{\rm U}(N)_{L,R} \cong {\rm
  U}(1)_{L,R}\otimes [{\rm SU}(N)/\mathbb{Z}(N)]_{L,R},
\label{unsun}
\end{equation}
and the group
${\rm U}(1)_L\otimes {\rm U}(1)_R$ is isomorphic to the group ${\rm
  U}(1)_V\otimes {\rm U}(1)_A$ of vector and axial ${\rm U}(1)$
transformations, the classical symmetry group of the theory can be
written as
\begin{equation}
{\rm U}(1)_V\otimes {\rm U}(1)_A\otimes [{\rm SU}(N_f)/\mathbb{Z}(N_f)]_L
                     \otimes [{\rm SU}(N_f)/\mathbb{Z}(N_f)]_R.
\label{totsym}
\end{equation}
The vector subgroup ${\rm U}(1)_V$ corresponds to the quark-number
conservation and it is not expected to play any role at the
transition.  The ${\rm U}(1)_A$ symmetry is broken to $\mathbb{Z}(N_f)_A$ 
by quantum fluctuations, since the divergence of the corresponding current
presents a quantum anomaly proportional to the topological charge
density.  This reduces the relevant symmetry to~\cite{PW-84} 
\begin{equation}
[{\rm SU}(N_f)_L\otimes {\rm SU}(N_f)_R]/\mathbb{Z}(N_f)_V. 
\label{relsymm}
\end{equation}
At zero temperature, the hadronic
spectrum shows that this symmetry is spontaneously broken to ${\rm
  SU}(N_f)_{V}$ with $N_f^2-1$ Goldstone particles (pions and kaons)
and a nonzero quark condensate $\langle \bar{\psi} \psi \rangle$.  The
large mass difference between the pseudoscalar flavor singlet and nonsinglet
mesons, such as $\eta$ and $\eta'$, which have the same quark content,
reflects the quantum breaking of the ${\rm U}(1)_A$ symmetry.

At finite temperature, a phase transition occurs at a critical
temperature $T_c$, $T_c\simeq 160$ MeV for $N_f=2$. Above $T_c$, 
chiral symmetry is restored and the quark condensate vanishes.
Therefore, the symmetry-breaking pattern at the chiral transition is
expected to be
\begin{equation}
[{\rm SU}(N_f)_L\otimes {\rm SU}(N_f)_R]/\mathbb{Z}(N_f)_V \rightarrow 
{\rm SU}(N_f)_V/\mathbb{Z}(N_f)_V,
\label{sym1}
\end{equation}
with a matrix-like order parameter given by the expectation value of
the quark bilinear $\Psi_{ij} \equiv \bar{\psi}_{L,i} \psi_{R,j}$.  In
the case of two flavors, i.e. $N_f=2$, the symmetry-breaking pattern
(\ref{sym1}) is equivalent to that of the O(4) vector model, i.e., to
O(4)$\to$O(3)
\cite{PV-02,GZ-98,Has-01,PPV-03,EFS-03,Deng-06,HV-11,EK-12,BJW-99}.  Thus, in
the case of a continuous transition, the critical behavior of the
model with two massless flavors is expected to belong to the
three-dimensional (3D) O(4) universality class.

The symmetry-breaking pattern at the transition significantly changes
if also the ${\rm U}(1)_A$ symmetry is restored.  The anomaly effects
breaking the ${\rm U}(1)_A$ symmetry are related to the topological
properties of QCD. Semiclassical instanton calculations predict a
substantial suppression of the instanton density for $T\gg T_c$, where
the dilute instanton gas (DIG) model is expected to provide a reliable
approximation~\cite{GPY-81}.  For example, in QCD with $N_f$ light
flavors of mass $m$, the topological susceptibility $\chi$ is expected
to decay asymptotically as~\cite{GPY-81}
\begin{equation}
\chi \sim m^{N_f} \, T^{-\kappa},  
\quad 
\kappa = {11\over 3}N_c + {1\over 3} N_f - 4 
\label{chisup}
\end{equation}
where $N_c$ is the number of colors.  For $N_c=3$ and $N_f=2$ we have
$\kappa=23/3$.  Although $\chi$ vanishes in the massless limit, the
Dirac zero modes associated with the instantons induce a residual
contribution to the ${\rm U}(1)_A$ symmetry breaking, giving rise to a
difference between the susceptibilities of the so-called $\pi$ and
$\delta$ channels at high $T$ \cite{Bazetal-12,CCCLMV-99},
which behaves as $\chi_\pi-\chi_\delta\sim T^{-\kappa}$ in the chiral
massless limit.

The breaking of the ${\rm U}(1)_A$ symmetry at finite $T$, and its
role at the chiral transition, has been much
investigated~\cite{Bazetal-12,CAFHKMN-13,Buchetal-13,AFT-12,
  Karsch-00,Vranas-99,CCCLMV-99,KLS-98,Bernard-etal-97,GR-13,MS-13,
  MM-13,Pawlowski-98,BCM-96,EHS-96,LH-96,Cohen-96,Shuryak-94}.  Monte
Carlo (MC) simulations of lattice
QCD~\cite{Bazetal-12,CAFHKMN-13,Buchetal-13,AFT-12,
  Karsch-00,Vranas-99,CCCLMV-99,KLS-98,Bernard-etal-97} find a
substantial suppression of the ${\rm U}(1)_A$ anomaly effects at large
$T$, as predicted by the DIG model.  These results are supported by
numerical investigations of pure ${\rm SU}(N)$ gauge theories, which
show that the topological susceptibility is rapidly suppressed above
the deconfinement transition, see, e.g., Ref.~\cite{VP-09} and
references therein, and that the DIG regime sets in quite early for
$T\gtrsim T_c$~\cite{BDPV-13}.  There are also some claims of an exact
restoration of the ${\rm U}(1)_A$ symmetry at the chiral
transition~\cite{AFT-12,CAFHKMN-13}.

It is thus worth investigating the nature of the finite-$T$ chiral
transition in the case the ${\rm U}(1)_A$ symmetry is effectively
restored, and the relevant symmetry-breaking pattern is
\begin{equation}
[{\rm U}(N_f)_L\otimes {\rm U}(N_f)_R]/{\rm U}(1)_V \rightarrow 
{\rm U}(N_f)_V/{\rm U}(1)_V, 
\label{sym2}
\end{equation}
instead of that reported in Eq.~(\ref{sym1}).

Up to now we have discussed the case of the model with $N_f$ massless
flavors.  However, in nature quarks have a finite mass. Since $u$ and
$d$ quarks are very light, one expects that the correct physical
behavior can be obtained by considering their masses as a perturbation
in the theory with $N_f=2$.  According to renormalization-group (RG)
theory, if the transition is continuous in the chiral massless limit,
then an analytic crossover is expected for nonzero values of the quark
masses $m_f$, because the quark masses act as external fields coupled
to the order parameter. Still, the presence of a close continuous
transition gives rise to scaling relations depending on the fermion
mass $m_f=m$ and on the reduced temperature $t\equiv (T-T_c)/T_c$. For
instance, the fermion condensate is expected to scale as
\begin{eqnarray}
\langle \bar\psi \psi \rangle \propto m^{1/\delta} E(m^{-1/(\beta+\gamma)} t),
\label{scalrel}
\end{eqnarray}
where $\delta$, $\beta$ and $\gamma$ are appropriate critical
exponents determined by the universality class of the transition, see
e.g. Refs.~\cite{WK-72,Fisher-74,ZJ-book,PV-02}, and $E(x)$ is
an universal scaling function (apart from trivial normalizations).  On
the other hand, a first-order transition is generally robust against
perturbations. Therefore, if the massless theory undergoes a
first-order transition, we expect a first-order transition also for
small nonvanishing values of the masses, up to an endpoint $m^*$,
around which a 3D Ising critical behavior is expected.  For larger
fermion masses the phase transition disappears and we have an analytic
crossover as well.

To make contact with experiments, it is also necessary to take into
account the massive strange quark $s$, whose mass ($m_s\approx 100$
MeV) is comparable with $T_c$.  Since the transition is expected to be
of first order for $N_f=3$ light degenerate quarks, we also expect a
first-order transition when increasing $m_s$ (keeping $m_u=m_d=0$), at
least for sufficiently small values of $m_s$.  For larger values of
$m_s$ there are two possibilities, depending on the nature of the
transition for $N_f=2$ degenerate quarks, corresponding to the limit
$m_s\to\infty$.  In one case we may have a first-order transition line
which extends for all values of $m_s$. Alternatively, the first-order
transition line extends up to a finite $m_s^*$, then the transition
becomes continuous for $m_s>m_s^*$, and in particular in the limit
$m_s\rightarrow \infty$; $m_s^*$ is a tricritical point, separating
the first-order transition line from the critical line, which implies
that the critical behavior for $m_s=m_s^*$ should be described by
mean-field theory, with logarithmic corrections.

The nature of the chiral transition has been extensively studied.  In
spite of several MC studies of different lattice QCD formulations with
two light quarks
~\cite{CP-PACS-01,MILC-00,KLP-01,KS-01,EHMS-01,DDP-05,KS-06,
  FP-07,BDDPS-12,Bazetal-12b,Burger-etal-13}, the nature of the chiral
transition is still controversial.  Some MC results favor a continuous
transition, but are not sufficiently accurate to clearly identify the
corresponding universality class.  Other MC studies report instead
evidence of a first-order transition.  For quark masses close to their
physical values, the results of MC
simulations~\cite{Be-etal-05,AFKS-06,
  Ch-etal-06,AEFKS-06,Bazetal-09,Aoki-etal-09,Cheng-etal-10,
  Borsanyi-etal-10} support a crossover scenario: the low-$T$ and
high-$T$ regimes are not separated by a phase transition, but rather
by a crossover region in which the thermodynamic quantities change
rapidly, but continuously, in a relatively narrow temperature
interval.

The universal features of the chiral transition can be investigated
within the RG framework~\cite{WK-72,Fisher-74}.  They are determined
by a few global properties, such as the space
dimensionality $d$ ($d=3$ for the finite-$T$ QCD transition),
the nature and the symmetry of the order parameter (a complex matrix
related to the bilinear quark operators $\bar{\psi}_{Li}\psi_{Rj}$),
and the symmetry-breaking pattern (which is given by Eqs.~(\ref{sym1})
or (\ref{sym2}) depending on the role played by the ${\rm U}(1)_A$
anomaly). For this purpose one considers the RG flow in the space of
Lagrangians which satisfy the above-reported general properties and
determines the fixed points (FPs) of the flow.  In the absence of a
stable FP, only first-order transitions between the disordered and
ordered phases are possible.  On the other hand, if a stable FP
exists, the transition may be continuous, and the usual critical
exponents $\nu$, $\eta$, etc... are related to the eigenvalues of the
linearized flow around the FP. However, it is important to stress
that, even in the presence of a stable FP, some systems may still
undergo a first-order transition.  From the RG point of view, this
occurs in systems which are not in the attraction domain of the stable
FP.

To determine the RG behavior of the model, one can use standard
perturbative field-theoretical approaches \cite{ZJ-book,PV-02}. The
first RG study of the effective model with symmetry-breaking pattern
(\ref{sym2}) was presented by Pisarski and Wilczek \cite{PW-84}, who
performed a one-loop calculation within the $\epsilon$ expansion,
$\epsilon = 4 -d$, finding no stable FP close to $d=4$. This result
suggests a first-order transition for the model with symmetry-breaking
pattern (\ref{sym2}).  However, subsequent analyses of the RG flow
directly in three dimensions, based on high-order perturbative
expansions (up to six loops), have provided the evidence of a stable
3D FP~\cite{BPV-05-l,BPV-05,BPV-03}.  Thus, the transition may also be
continuous if the ${\rm U}(1)_A$ symmetry is effectively restored at
$T_c$. However, its universality class differs from the 3D ${\rm
  O}(4)$ universality class.

Table~\ref{summarytqcd} summarizes the results of these RG analyses,
reporting the possible transitions for various values of $N_f$, and
for the two symmetry breaking patterns (\ref{sym1}) and (\ref{sym2}).
When a continuous transition is possible, the corresponding
universality class is reported.

\begin{table}[tbp]
  \caption{Summary of the RG predictions for the finite-$T$
    QCD transition, as a function of the number $N_f$ of light
    flavors. We distinguish two cases, depending whether the U(1)$_A$
    symmetry is broken or effectively restored. When
    a continuous transition is possible, we specify the corresponding
    3D universality class by reporting its symmetry-breaking
    pattern.}
\label{summarytqcd}
\begin{ruledtabular}
\begin{tabular}{ccc}
\multicolumn{1}{c}{$N_f$}& 
\multicolumn{1}{c}{${\rm U}(1)_A$ broken}& 
\multicolumn{1}{c}{${\rm U}(1)_A$ restored}\\
\hline
$1$ & crossover or 1$^{\rm st}$ord  & ${\rm O}(2)\to\mathbb{Z}_2$ or 
1$^{\rm st}$ord  \\

$2$ & ${\rm O}(4)\to{\rm O}(3)$ or 1$^{\rm st}$ord& 
${\rm U}(2)_L\otimes {\rm U}(2)_R\to{\rm U}(2)_V$   \\
&& or 1$^{\rm st}$ord   \\
$\ge 3$ & 1$^{\rm st}$ord  & 1$^{\rm st}$ord  \\
\end{tabular}
\end{ruledtabular}
\end{table}

In this paper we extend previous field-theoretical
studies~\cite{BPV-05-l,BPV-05,BPV-03} of the 3D ${\rm U}(2)_L \otimes
{\rm U}(2)_R\to {\rm U}(2)_V$ universality class. Our purpose is to
provide accurate predictions for the critical features of this
universality class, for which we know much less compared with the
${\rm O}(N)$ vector universality classes.  We study the RG flow of the
multiparameter $\Phi^4$ theory with the same symmetry breaking.  For
this purpose, we consider two different 3D perturbative schemes: 
the massive zero-momentum (MZM) scheme 
\cite{Parisi-80,ZJ-book,PV-02} and the 3D minimal subtraction 
scheme ${\overline {\rm MS}}$ without $\epsilon$ expansion
\cite{Dohm-85,SD-89}.

The resummation of the 
perturbative expansions of the $\beta$ functions (known to six and five loops 
in the MZM and ${\overline {\rm MS}}$ scheme, respectively)
allows us to compute the
RG trajectories starting from the unstable Gaussian FP of the free
theory. They approach a stable FP in both schemes for an extended
region of bare parameters.  Moreover, we estimate the critical
exponents by computing the expansions of appropriate RG
functions and evaluating them at the stable FP.

The paper is organized as follows.  In Sec.~\ref{secrgarg} we review
the universality and RG arguments which we use to investigate the
finite-$T$ transition in QCD with $N_f=2$ light flavors. In
particular, we define the effective theory that is relevant for the
model with symmetry-breaking pattern (\ref{sym2}). In
Sec.~\ref{rgflux} we study the RG flow in the space of the
renormalized couplings; in particular we determine the RG trajectories
that start at the unstable Gaussian FP of the free quadratic theory 
and flow towards the 
stable FP controlling the critical behavior at the transition.
Moreover, we determine the critical exponents by evaluating
appropriate RG functions at the stable FP. Finally, in
Sec.~\ref{conclusions} we draw our conclusions. In App.~\ref{MSseries}
we report the perturbative expansions used in the paper to
determine the RG flow and the critical exponents.

\section{RG analysis of the chiral transition} 
\label{secrgarg}

The nature of the finite-$T$ chiral transition in QCD can
be investigated using universality and RG
arguments~\cite{PW-84,BPV-05-l,BPV-03,BPV-05,v-07}.  In this section
we review them, focussing on the finite-$T$ transition in QCD
with two light flavors.

Let us first assume that the phase transition at $T_c$ is continuous
for vanishing quark masses.  In this case the length scale of the
critical modes diverges approaching $T_c$, becoming eventually much
larger than $1/T_c$, which is the size of the euclidean ``temporal''
dimension at $T_c$.  Therefore, the asymptotic critical behavior is
associated with a 3D universality class with the same symmetry
breaking pattern and the order parameter is an $N\times N$
complex-matrix field $\Phi_{ij}$, related to the bilinear quark
operators $\bar{\psi}_{Li}\psi_{Rj}$.  Nonvanishing quark masses can
be accounted for by an external field coupled to the order
parameter.

To determine the critical behavior we consider the most general 
Landau-Ginzburg-Wilson (LGW) $\Phi^4$ theory compatible with the given
symmetry breaking. If Eq.~(\ref{sym2}) holds, the theory is given by
\begin{eqnarray}
{\cal L}_{{\rm U}(N)} &=& {\rm Tr} (\partial_\mu \Phi^\dagger) (\partial_\mu \Phi)
+r {\rm Tr} \Phi^\dagger \Phi \label{LUN}\\
&&+ {u_0\over 4} \left( {\rm Tr} \Phi^\dagger \Phi \right)^2
+ {v_0\over 4} {\rm Tr} \left( \Phi^\dagger \Phi \right)^2 ,
\nonumber
\end{eqnarray}
where the field $\Phi_{ij}$ is a generic $N\times N$ complex
matrix. The symmetry is U$(N)_L\otimes$U$(N)_R$, which breaks to
U$(N)_V$ if $v_0>0$, thus providing the LGW theory relevant for QCD
with two light flavors.  The reduction of the symmetry to
SU$(N)_L\otimes$SU$(N)_R$ for QCD, due to the axial anomaly, can be
achieved by adding additional quadratic and quartic terms containing
the determinant of the field $\Phi$~\cite{BPV-03}.  We return to this
point later.

The critical behavior at a continuous transition is controlled by the
FPs of the RG flow, which are determined by the common zeroes of the
$\beta$-functions associated with the quartic parameters.  To study
the RG flow of the $\Phi^4$ theory (\ref{LUN}), we consider two
different perturbative schemes. 

In the massive zero-momentum (MZM)
scheme~\cite{Parisi-80,ZJ-book,PV-02} one performs the perturbative
expansion directly in three dimensions, in the critical region of the
disordered phase.  The MZM perturbative expansions of the 
$\beta$ functions and of the critical exponents  have been computed to
six loops, requiring the computation of approximately 1000 Feynman
diagrams. The six-loop series of the $\beta$ functions were reported
in Ref.~\cite{BPV-03}, here we also report those of the RG
functions associated with the critical exponents, in
App.~\ref{sermzm}.

In the 3D ${\overline {\rm MS}}$ scheme one considers the
massless critical theory: one uses dimensional regularization and the
modified minimal-subtraction prescription, thus the RG functions are
obtained from the divergences appearing in the perturbative expansion
of the correlation functions~\cite{tHV-72}.  In the standard
$\epsilon$-expansion scheme~\cite{WF-72}, the FPs, i.e., the common
zeroes of the $\beta$-functions, are determined perturbatively as
expansions in powers of $\epsilon\equiv d-4$, while exponents are
obtained by expanding the corresponding RG functions computed at the
FP in powers of $\epsilon$.  
Physical results are then obtained by extrapolating the results to 
$d=3$.  This procedure assumes the
existence of a FP for $\epsilon\to 0$, i.e., close to four dimensions. 
Therefore, it allows one to
determine only those three-dimensional FPs which can be defined, 
by analytic continuation, close to four dimensions. 
Other FPs, which do not have a four-dimensional counterpart, cannot 
be detected.  This problem is overcome by the 3D
$\overline{\rm MS}$ scheme without $\epsilon$
expansion~\cite{Dohm-85,SD-89,CPPV-04}. The RG functions $\beta_{u,v}$
and $\eta_{\phi,t}$ are the $\overline{\rm MS}$ functions.  However,
$\epsilon\equiv 4-d$ is no longer considered as a small quantity, but
it is set equal to its physical value ($\epsilon=1$ in our case)
before computing the FPs.  This provides a well defined 3D
perturbative scheme which allows us to compute universal quantities,
without the need of expanding around $d=4$~\cite{Dohm-85,SD-89}.  In
the $\overline{\rm MS}$ scheme the RG series of the $\beta$ functions
are known up to five loops~\cite{CP-04}; here we also present the
five-loop series of the RG functions associated with the critical
exponents. They are reported in App.~\ref{serms}.

The physically relevant results are obtained by resumming the
perturbative expansions (which are divergent but Borel summable), using methods 
that take into account their large-order behavior, which
is computed by semiclassical (hence, intrinsically nonperturbative)
instanton calculations \cite{LZ-77,ZJ-book,CPV-00}. For the model (\ref{LUN})
the large-order behavior is discussed in Refs.~\cite{BPV-03,CPPV-04}.
The method we use is described in Refs.~\cite{ZJ-book,CPV-00}. 
Resummations depend on two parameters, which are optimized in the procedure
~\cite{footnoteresum}.

\section{RG perturbative results for the 3D U(2)$\otimes$U(2) theory}
\label{rgflux}

\begin{figure}[tbp]
\includegraphics*[scale=\graphicscale]{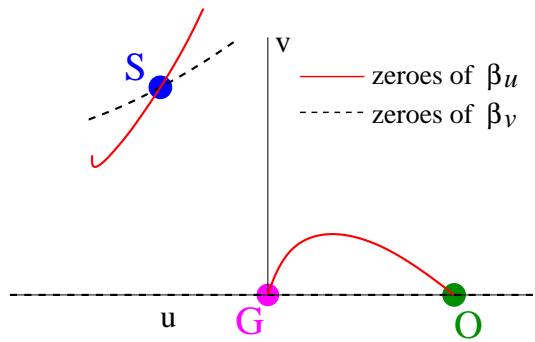}
\caption{Zeroes of the $\beta$-functions $\beta_u$ and $\beta_v$
  associated with the quartic couplings of the Lagrangian
  (\protect\ref{LUN}) for $N=2$.  For $v\ge 0$ the $\beta$-functions
  have three common zeroes, corresponding to three FPs: the Gaussian
  (G) and O(8) (O) FPs along the $v=0$ axis are unstable, while the FP
  (S) with $v>0$ and $u<0$ is stable.  }
\label{zeroesu22}
\end{figure}

In this section we study the 3D RG flow of model (\ref{LUN}) in
the case relevant for QCD with two light flavors, i.e.  for $N=2$ and
$v_0>0$. Fig.~\ref{zeroesu22} provides a sketch of the locations of
the stable and unstable FPs.  The RG trajectories,
starting from the unstable Gaussian FP (denoted by $G$ in
Fig.~\ref{zeroesu22}) of the quadratic theory, flow toward a
nontrivial FP (denoted by $S$) for an extended region of quartic bare
parameters $v_0$ and $u_0$, which implies the stability of the
FP.  We then determine the critical exponents at the stable FP.

\subsection{RG trajectories toward the stable FP}
\label{rgfp}

We first consider the MZM scheme, where one expands in powers of the
zero-momentum renormalized quartic couplings.  The theory is
renormalized by introducing a set of zero-momentum conditions for the
one-particle irreducible two-point and four-point correlation
functions of the 2$\times$2 matrix-like field $\Phi_{ab}$:
\begin{eqnarray}
&&\Gamma^{(2)}_{a_1a_2,b_1b_2}(p) = 
\delta_{a_1b_1}\delta_{a_2b_2} Z_\phi^{-1} \left[ m^2+p^2+O(p^4)\right],
\quad \label{ren1}  \\
&&\Gamma^{(4)}_{a_1a_2,b_1b_2,c_1c_2,d_1d_2}(0) 
= 2 \pi Z_\phi^{-2} m^{4-d} \times
\label{ren2}  \\
&&
\quad \times \left(u U_{a_1a_2,b_1b_2,c_1c_2,d_1d_2} +  
v V_{a_1a_2,b_1b_2,c_1c_2,d_1d_2}\right),
\nonumber 
\end{eqnarray}
where $Z_\phi$ is the renormalization constant of the order-parameter
field $\Phi$, and $U,\,V$ are appropriate form factors defined so that
$u\propto u_0/m$ and $v\propto v_0/m$ at the leading tree order (more
details are reported in Ref.~\cite{BPV-03}, where the coupling
$u,v$ were denoted by $\bar{u},\,\bar{v}$).  The FPs of the theory
are given by the common zeroes of the Callan-Symanzik
$\beta$-functions
\begin{equation}
\beta_u(u,v) = \left. m{\partial u\over \partial m}\right|_{u_0,v_0},\quad
\beta_v(u,v) = \left. m{\partial v\over \partial m}\right|_{u_0,v_0}.
\label{betaf}
\end{equation}
The resummation of the six-loop series of the $\beta$ functions, as
outlined in Refs.~\cite{BPV-03,footnoteBPV03}, finds a FP (point
$S$ in Fig.~\ref{zeroesu22}) at~\cite{BPV-05,footnote-err}
\begin{equation}
u^* = -3.4(3), \qquad  v^* = 5.3(3),
\label{MZMFP}
\end{equation}
beside the unstable Gaussian FP at $u=v=0$ and the O(8) FP along the
$v=0$ axis, see Fig.~\ref{zeroesu22}.  
A FP is stable if all eigenvalues of the corresponding
stability matrix, $\Omega_{ij} = {\partial \beta_i/\partial g_j}$
(where $g_{1,2}$ corresponds to $u,v$) have positive real part.  The
numerical analysis of the stability matrix at the FP (\ref{MZMFP})
favours its stability~\cite{BPV-05-l}.  In the following we provide a
more direct evidence of the stability of this FP showing that the RG
trajectories in the space of the renormalized couplings flow towards 
this FP for an extended region of bare parameters.  Physically, the
values $u^*$ and $v^*$ at a stable FP are the two independent (RG
invariant) couplings which describe the zero-momentum behavior of the
quartic correlations in the critical region of the disordered
phase~\cite{PV-02}.

The existence of a stable FP is confirmed by the analysis of the
$\beta$ functions in the 3D $\overline{\rm MS}$ scheme.  The
renormalized couplings are again defined from the irreducible
four-point correlation function, and the ${\overline {\rm MS}}$
$\beta$ functions are 
\begin{equation}
\beta_u(u,v) = \left. \mu{\partial u\over \partial \mu}\right|_{u_0,v_0},\quad
\beta_v(u,v) = \left. \mu{\partial v\over \partial \mu}\right|_{u_0,v_0},
\label{betafms}
\end{equation}
where $\mu$ is the energy scale of this massless scheme.  See
Ref.~\cite{CP-04} for more details. In the 3D ${\overline {\rm
    MS}}$ scheme we set $\epsilon=4-d=1$ and then resum the series
using Borel resummation techniques.  The five-loop series of
the $\beta$ functions are reported in App.~\ref{serms}.  Again a
nontrivial common zero of the $\beta$ functions is found at~\cite{footnote-err}
\begin{equation}
u^* = -0.55(6), \qquad v^* = 1.22(9),
\label{MSFP}
\end{equation}
which is represented by the point $S$ in Fig.~\ref{zeroesu22}.  Note
that the renormalized couplings of the MZM and ${\overline {\rm MS}}$
perturbative schemes correspond to different quartic couplings, thus
their FP values, cf. Eqs.~(\ref{MZMFP}) and (\ref{MSFP}), differ.

In order to check the stability of these FPs and determine their
attraction domain, we study the RG flow in the space of the
renormalized parameters. If $u$ and $v$ are the renormalized couplings
and $u_0$, $v_0$ the corresponding Lagrangian couplings that satisfy
$u\approx u_0/m$ and $v\approx v_0/m$ at tree level ($m$ is the
zero-momentum mass in the MZM scheme and the renormalization energy
scale $\mu$ in the $\overline{\rm MS}$ scheme), the RG trajectories
are determined by solving the differential equations
\begin{eqnarray}
&&-\lambda {d u\over d\lambda} = \beta_u(u(\lambda),v(\lambda)),\nonumber\\
&&-\lambda {d v\over d\lambda} = \beta_v(u(\lambda),v(\lambda)),
\label{rgfloweq}
\end{eqnarray}
where $\lambda\in [0,\infty)$, with the initial conditions
\begin{eqnarray} 
&&u(0) = v(0) = 0 ,\nonumber \\
&& \left. {d u\over d\lambda} \right|_{\lambda=0} = s\equiv {u_0\over v_0},\qquad
\left. {d v\over d\lambda} \right|_{\lambda=0} = 1, \label{ini-rgflow}
\end{eqnarray}
where $s$ parametrizes the different RG trajectories in terms of the
bare quartic parameters.  Note that the initial condition $dv/d\lambda
= +1$ for $v$ is required by the theory. Indeed, systems with $v_0 <
0$ are associated with transitions with a different symmetry-breaking
pattern, i.e.
\begin{equation}
{\rm U}(N_f)_L\otimes {\rm U}(N_f)_R \to {\rm U}(N_f-1)_L\otimes {\rm U}(N_f-1)_R.
\label{otherSB}
\end{equation}

Since the stability of the $\Phi^4$ theory (\ref{LUN}) requires~\cite{BPV-03} 
\begin{equation}
u_0 + v_0 > 0,\qquad  u_0 + {1\over 2} v_0 > 0,
\label{stabcond}
\end{equation}
physical systems corresponding to the effective theory (\ref{LUN})
with $v_0>0$ and $s < -1/2$ are expected to undergo a first-order
phase transition.

\begin{figure}[tbp]
\includegraphics*[scale=\graphicscale]{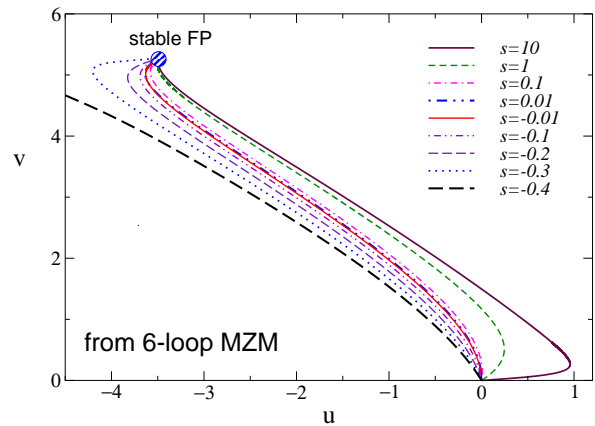}
\caption{The RG flow in the renormalized coupling space of the
  MZM scheme, for several values of the ratio 
$s\equiv u_0/v_0$ of the bare 
quartic parameters.}
\label{mzmrgflow}
\end{figure}

\begin{figure}[tbp]
\includegraphics*[scale=\graphicscale]{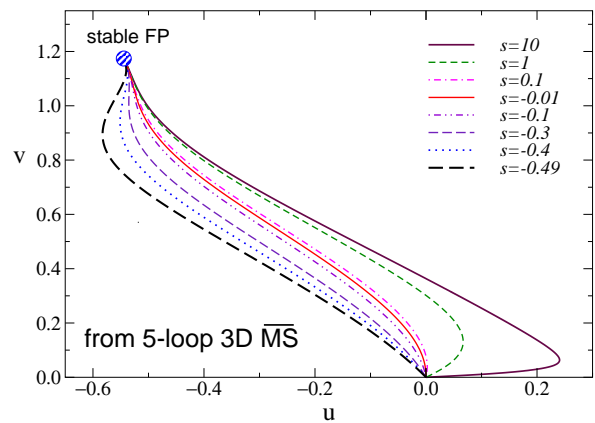}
\caption{ 
RG flow in the renormalized coupling space of the
  massless 3D ${\overline {\rm MS}}$ scheme,
for several values of the ratio 
$s\equiv u_0/v_0$ of the bare 
quartic parameters.}
\label{msrgflow}
\end{figure}

The RG trajectories for $s>-1/2$ are determined by solving
Eq.~(\ref{rgfloweq}) after resumming the expansions of the
$\beta$ functions.  In Figs.~\ref{mzmrgflow} and \ref{msrgflow} we
report the RG flow for several values of the ratio $s$, as obtained by
a particular choice of the approximants that are used to perform the
resummation of the perturbative $\beta$
functions~\cite{CPV-00,ZJ-book,footnoteresum}.  Different approximants
show analogous qualitative behaviors when they are chosen in the 
optimal region, defined as outlined in Refs.~\cite{CPV-00,ZJ-book}.  The RG
trajectories in both schemes are attracted by a FP for an extended
region of bare quartic parameters $u_0,\,v_0$, which implies that the
FP is stable. In the $\overline{\rm MS}$ scheme, all trajectories with
$s \gtrsim -0.5$ flow towards the FP given in Eq.~(\ref{MZMFP}), which
is indeed the stable FP of the model.  In the MZM scheme, we find the
same as long as $s \gtrsim -0.4$.  The trajectory that corresponds to
$s = -0.4$ runs away, into the region in which the perturbative series
can no longer be resummed (the closest Borel singularity is on the
positive real axis), hence we are not able to determine its
large-$\lambda$ behavior. In any case, both perturbative schemes show
the presence of a stable FP. Moreover, both schemes consistently find
that the attraction domain of the bare quartic parameters corresponds 
to $s = u_0/v_0 \gtrsim -0.5$.

Note that no stable FP is found close to $d=4$, in agreement with the
one-loop $\epsilon$-expansion calculation of Ref.~\cite{PW-84}, see
also Ref.~\cite{CP-04}. However, the extension of this result to the
relevant $d=3$ dimension fails.  This is not the only physically
interesting case in which $\epsilon$-expansion calculations
fail to provide the correct physical picture in three dimensions. For
example, this also occurs for the Ginzburg-Landau model of
superconductors, in which a complex scalar field couples to a gauge
field: although $\epsilon$-expansion calculations do not find a stable
FP \cite{HLM-74}, thus predicting first-order transitions, it is now
well established (see, e.g., Refs.~\cite{KKL-98,MHS-02}) that 3D
systems described by the Ginzburg-Landau model can also undergo a
continuous transition---this implies the presence of a stable FP in
the 3D Ginzburg-Landau theory---in agreement with
experiments~\cite{GN-94}.  Other examples are provided by the LGW
$\Phi^4$ theories describing frustrated spin models with noncollinear
order~\cite{CPPV-04} and the $^3$He superfluid transition form the
normal to the planar phase~\cite{DPV-04}.  

Finally, we also mention that the RG flow of the U(2)$\otimes$U(2)
scalar theory has been also studied by methods based on approximate
solutions of functional RG equations.~\cite{BW-97,FKK-11} They have
not found evidence of a stable FP, but they were limited to
approximations keeping only the first terms of the derivative
expansion of the effective action.

\subsection{The critical exponents of the U(2)$\otimes$U(2) $\Phi^4$ theory.}
\label{rgcrexp}

We now compute the critical exponents by evaluating the corresponding RG
functions at the stable FP. In the MZM scheme they are given by
\begin{eqnarray} 
\eta_\phi(u,v) = {\partial \ln Z_\phi\over \partial \ln m}, \quad
\eta_t(u,v) = {\partial \ln Z_t \over \partial \ln m}, 
\label{etapt}
\end{eqnarray}
where $Z_\phi$ and $Z_t$ are the renormalization functions of the
field $\Phi$ and of the quadratic operator ${\rm Tr} \,\Phi^\dagger \Phi$,
respectively~\cite{BPV-03}. The six-loop series are reported
in App.~\ref{sermzm}.  We have performed an analogous calculation
in the ${\overline {\rm MS}}$ scheme. The perturbative five-loop series 
are reported in  App.~\ref{serms}.

The critical exponents are obtained by evaluating the resummed 
RG functions at the stable FP. In particular,
\begin{eqnarray}
\eta = \eta_\phi(u^*,v^*),\quad 
\nu = [2 - \eta + \eta_t(u^*,v^*)]^{-1}.
\label{exprel}
\end{eqnarray}
Resumming the perturbative series by using the conformal-Borel method
\cite{ZJ-book}, we obtain
\begin{eqnarray}
&\nu=0.71(7), \quad  &\eta=0.12(1),
\quad [{\rm 6\,loop}\;{\rm MZM}],
\label{mzmest}\\
&\nu=0.76(10), \quad  &\eta=0.11(6),\quad
[{\rm 5\,loop}\;{\overline {\rm MS}}].
\label{msest}
\end{eqnarray}
The errors take into account the uncertainty on the location of the
FP, the dependence of the results on the resummation parameters
and the stability of the estimates with 
respect to the number of terms in the series \cite{footnote-err}.

Estimates (\ref{mzmest}) and (\ref{msest}) 
obtained in the two perturbative schemes are fully
consistent. This agreement provides a nontrivial crosscheck of the
accuracy of the analysis of the MZM and 3D ${\overline{\rm MS}}$
perturbative series: the resummation of the perturbative 
expansions in two different schemes give consistent results for
the universal quantities.  This fact may be hardly explained as an 
artefact of the resummation; it should instead be considered as a
robust evidence of the existence of a stable FP describing 3D
continuous transitions characterized by the symmetry-breaking pattern
U(2)$\otimes$U(2)$\to$U(2).

\section{Conclusions}
\label{conclusions}

In this paper we report a detailed study of the RG flow of the
$\Phi^4$ model (\ref{LUN}) for $N=2$, which is relevant for the
finite-$T$ chiral transition of two-flavor QCD if the ${\rm U}(1)_A$
symmetry is restored at the chiral transition.  For this purpose we
consider two field-theoretical perturbative schemes: the MZM scheme, 
defined in the disordered massive phase, and the 
3D ${\overline {\rm MS}}$ scheme without $\epsilon$ expansion, which 
considers the massless critical theory.
Extending
previous RG studies~\cite{BPV-05-l,BPV-05,BPV-03}, we verify the
existence of a stable FP with $v_0> 0$, which is the relevant domain
for the symmetry-breaking pattern (\ref{sym2}).  We study the RG flow
in the space of the renormalized quartic couplings, as obtained by the
analysis of the perturbative expansions of the $\beta$ functions,
computed to six and five loops in the MZM and 3D ${\overline
  {\rm MS}}$ scheme, respectively.  In both cases the RG trajectories starting
from the unstable Gaussian FP flow towards a nontrivial stable FP 
for an extended region of
the bare quartic parameters $u_0,\,v_0$, i.e. for $u_0/v_0\gtrsim
-0.5$.  This implies that systems corresponding to an effective
Lagrangian with $u_0/v_0 \gtrsim - 0.5$ undergo a continuous transition.  We
also estimate the corresponding critical exponents, obtaining
consistent results in the two field-theoretical schemes considered,
cf. Eqs.~(\ref{mzmest}) and (\ref{msest}).  On the other hand, systems
corresponding to $u_0/v_0 \lesssim -0.5$ are expected to undergo a
first-order transition.

The existence of a stable FP with $v > 0$ implies that the 
finite-$T$ chiral transition of two-flavor QCD
can be continuous also if the ${\rm U}(1)_A$ symmetry is effectively
restored at $T_c$. Although the critical behavior differs from that
expected in the case of a substantial ${\rm U}(1)_A$ symmetry breaking
around $T_c$, which is the 3D ${\rm O}(4)$ vector universality class,
we note that differences are small.  For instance, the critical exponents of the
${\rm O}(4)$ universality class~\cite{Has-01,HV-11}, $\nu=0.749(2)$,
$\eta=0.0365(10)$, $\delta=(5-\eta)/(1+\eta) \approx 4.789(6)$,
$\beta=\nu(1+\eta)/2\approx 0.388(1)$, $\gamma=\nu(2-\eta)=1.471(4)$
and $\alpha=2-3\nu=-0.247(6)$ are close to those we have obtained for
model (\ref{LUN}). In the MZM scheme we obtain $\nu = 0.71(7)$, $\eta
= 0.12(1)$, $\delta = 4.3(1)$, $\beta = 0.40(4)$, $\gamma = 1.3(1)$,
and $\alpha = 0.1(2)$. In the $\overline{\rm MS}$ scheme we obtain
instead $\nu = 0.76(10)$, $\eta = 0.11(6)$, $\delta = 4.4(3)$, $\beta
= 0.42(6)$, $\gamma = 1.4(2)$, and $\alpha = -0.3(3)$.  Thus,
only very accurate estimates of the  critical exponents can distinguish the 
two different critical behaviors.

We stress that the existence of a universality class does not exclude
that some systems with the same order parameter and symmetry-breaking
pattern undergo a first-order transition.  This occurs when the system
is outside the attraction domain of the stable FP, i.e., when the
system at the transition is effectively described by the Lagrangian
(\ref{LUN}) with quartic parameters $u_0$ and $v_0$ belonging to the
large region $u_0/v_0\lesssim -0.5$. The nature (first-order or
continuous) of the transition is nonuniversal, since it depends on the
details of the model and not only on the global features that
characterize the universality class.  For example, the model
considered in Ref.~\cite{CM-07}, which corresponds to two-flavor lattice QED
with the same symmetry breaking ${\rm U}(2)_L\otimes {\rm
  U}(2)_R\rightarrow {\rm U}(2)_V$, shows a first-order
transition. This result cannot be extended to all transitions with the
same symmetry breaking, because it is indeed possible that this
model corresponds to a run-away RG trajectory, while the case relevant
to QCD may belong to the attraction domain of the stable FP, thus
undergoing a continuous transition.

\begin{figure}[tbp]
\includegraphics*[width=9truecm]{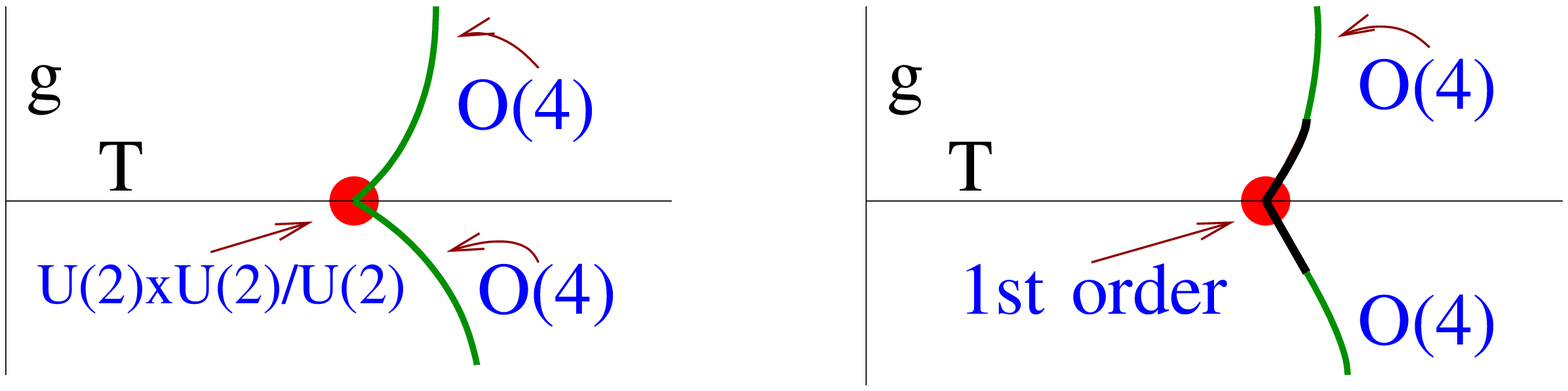}
\caption{Possible phase diagrams in the $T$-$g$ plane for the effective model
  with symmetry-breaking pattern (\ref{sym1}). The parameter $g$ is 
  proportional to the U(1)$_A$ symmetry breaking, hence for $g=0$
  we reobtain the model with symmetry-breaking pattern (\ref{sym2}). 
  On the left panel, the multicritical transition at $g=0$ is continuous,
  on the right panel it is of first order. Thick black 
  lines indicate first-order transitions. The endpoints of 
  the first-order transition lines correspond to mean-field 
  transitions with logarithmic corrections.  }
\label{pd}
\end{figure}

The results for model (\ref{LUN}) are also of interest if chiral
symmetry is not exactly restored at $T_c$, but U(1)$_A$ breaking
effects are small.  In this case we can parametrize the effective
Lagrangian as 
\begin{eqnarray}
&&{\cal L}_{{\rm SU}(2)} =
{\cal L}_{{\rm U}(2)}  
+ w_0 \left( {\rm det} \Phi^\dagger + {\rm det} \Phi \right)+
\label{LSU2}\\
&& {x_0\over 4} \left( {\rm Tr} \Phi^\dagger \Phi \right) 
         \left( {\rm det} \Phi^\dagger + {\rm det} \Phi \right) 
+ {y_0\over 4} \left[ ({\rm det} \Phi^\dagger)^2 + ({\rm det} \Phi)^2 \right],
\nonumber
\end{eqnarray}
where we added all terms up to dimension four which contain the
determinant of the $2\times 2$ order parameter field and leave a
residual SU(2)$\otimes$SU(2) symmetry.  In the context of 
two-flavor QCD, we may assume that $w_0,x_0,y_0\sim g$ where $g$
parametrizes the effective breaking of the U(1)$_A$ symmetry.  In the
$T$-$g$ plane, the U(2)$_L$$\otimes$U(2)$_R$ transition point becomes
a multicritical point \cite{BPV-03}, as Lagrangian (\ref{LSU2}) contains
two quadratic terms.  In Fig.~\ref{pd} we show two 
possible phase diagrams, depending on the nature of the transition at $g=0$.
In the first case (left panel of
Fig.~\ref{pd}), the transition is always continuous for $g\ne 0$ along
the critical line $T_c(g)$.  But if $|g|$ is small, we may observe a
crossover behavior controlled by the U(2)$\otimes$U(2) multicritical
point at $g=0$: the free energy should behave as ${\cal F}_{\rm sing}
\approx t^{3\nu} f(g t^{-\phi})$, where $t\propto T-T_c(g=0)$,
$\nu\approx 0.7$, and $\phi\approx 1.3$~\cite{footnotecpv}. In
practice, if $g$ is small, one might observe two different behaviors
depending on the distance of $T$ from the critical line $T_c(g)$. For
$|T-T_c(g)|$ not too small, the RG flow is influenced by the
U(2)$\otimes$U(2) multicritical point, hence one would observe an
effective critical behavior analogous to that for $g=0$. As $T$
approaches $T_c(g)$, this crossover behavior disappears and the O(4)
behavior is eventually observed.  In the other case, shown in the right
panel of Fig.~\ref{pd}, we expect first-order transitions to occur
also for $|g|$ small, up to an endpoint $g^*$ where mean-field
behavior with logarithmic corrections should be observed; then the
continuous transition for larger $g>g^*$ (note that $g^*$ does not
need to be small) is expected to belong to the O(4) universality
class. The available numerical MC results for QCD with two light
flavors do not yet allow us to distinguish between the above
scenarios.

Finally, we would like to discuss the possible scenarios for the
finite-$T$ transitions of the QCD-like theory with a large number $N_c$ of
colors, widening the parameter space to get further hints for the
relevant $N_c=3$ case.  We first note that the universality arguments
based on the global flavor symmetries do not depend on the number of
colors, thus they hold for any $N_c>3$ (keeping $N_f$ fixed) as well,
including $N_c\to\infty$.  At large $N_c$, keeping the number $N_f$ of
flavors fixed, we expect a first-order transition corresponding to the
deconfinement transition of pure ${\rm SU}(N_c)$ gauge theories for a
large number of colors, thus at~\cite{LTW-04,LRR-12}
$T_c/\sqrt{\sigma} = 0.545(2)+ O(N_c^{-2})$ where $\sigma$ is the
string tension.  The presence of $N_f=2$ fermion flavors, which
contribute to $O(1/N_c)$ according to standard large-$N_c$ scaling
arguments~\cite{Hooft-74}, cannot smooth out this transition whose
latent heat is $O(N_c^2)$~\cite{MP-07,LP-13,HY-12}.
This opens the road to other possible scenarios with respect to the
standard three-color QCD. Indeed, the chiral symmetry of the fermions
may be restored at the same transition point, or we may have another
transition at a larger temperature~\cite{HY-12}, like the case of QCD
with adjoint fermions~\cite{KL-99,EHS-05}, where the deconfinement and
chiral transitions occur at different temperatures.
Moreover, since the ${\rm U}(1)_A$ anomaly is suppressed by $1/N_c$ in
the large-$N_c$ limit~\cite{Witten-79}, the ${\rm U}(1)_A$ symmetry
breaking is further suppressed at large $N_c$, as also shown by the
behavior $\chi_\pi-\chi_\delta\sim T^{-\kappa}$, where the exponent
$\kappa\sim N_c$, cf. Eq.~(\ref{chisup}).  Therefore, with increasing
$N_c$, the effective symmetry breaking at the chiral transition with
two light flavors should be better and better described by ${\rm
  U}(2)_L\otimes {\rm U}(2)_R\rightarrow {\rm U}(2)_V$, rather than
${\rm SU}(2)_L\otimes {\rm SU}(2)_R \rightarrow {\rm SU}(2)_V$.

\medskip

{\em Acknowledgements.} We thank Claudio Bonati and Massimo D'Elia
for useful discussions.

\bigskip

\appendix

\section{High-order pertubative series of the ${\rm U}(2)\otimes {\rm U}(2)$
$\Phi^4$ theory}
\label{MSseries}

In this appendix we report the perturbative series used in
the paper to analyze the RG flow of the U(2)$\otimes$U(2) $\Phi^4$
theory (\ref{LUN}) and estimate the critical exponents, i.e. the
six-loop series of the MZM scheme and the five-loop series of the 3D
${\overline {\rm MS}}$ scheme.

\begin{widetext}

\subsection{The MZM series up to six loops}
\label{sermzm}

The $\beta$ functions of the MZM perturbative schemes 
have been already reported to six loops in Ref.~\cite{BPV-03}.
Here we report the six-loop RG functions defined in Eq.~(\ref{etapt}),
which allow us to evaluate the critical exponents through
Eqs.~(\ref{exprel}). 
They are given by
\begin{eqnarray}
&&\eta_\phi(u,v)  =   0.011574074 u^2 + 0.000964218 u^3 + 0.001280763 u^4 - 
    0.000212863 u^5 + 0.000487251 u^6   \label{etaMZM}\\
&& +  0.018518518 u v + 0.002314125 u^2 v + 0.004098443 u^3 v -
    0.00085145 u^4 v + 0.00233880 u^5  v  \nonumber \\
&& + 0.011574074 v^2  + 0.002241809 u v^2  + 
0.00595656 u^2 v^2 - 0.00169549 u^3 v^2 +    0.00540470 u^4 v^2  \nonumber \\
&& + 0.000771375 v^3 + 
0.00417400 u v^3 - 0.00183407 u^2 v^3 +  0.0071929 u^3 v^3 \nonumber\\
&& + 0.00105250 v^4 + 
0.000956991 u v^4 + 0.00550762 u^2 v^4 - 
0.000180727 v^5 + 0.00223754 u v^5 + 0.000372148 v^6,
\nonumber
\end{eqnarray}
and
\begin{eqnarray}
&&\eta_t(u,v) = -0.625 u + 0.078125 u^2 - 0.053818796 u^3 + 0.0282218 u^4 -
0.0265992 u^5 + 0.0230998 u^6 
\label{etatMZM}\\
&& -0.5 v + 0.125 u v - 0.12916511 u^2 v + 0.0903098 u^3 v
- 0.10639677 u^4 v + 0.11087907 u^5 v \nonumber \\
&& + 0.078125 v^2 - 0.12444547 u v^2 + 0.13414445 u^2 v^2 - 0.19715907 u^3 v^2 
+ 0.2561752 u^4 v^2 \nonumber \\
&& - 0.0430550 v^3 + 0.0964956 u v^3 - 0.1970603 u^2 v^3 +  0.3409476 u^3 v^3 
\nonumber \\
&& + 0.0241934 v^4 - 0.0993356 uv^4 + 0.260895 u^2 v^4 - 0.0193974 v^5 
+ 0.1057591 u v^5  + 0.017517 v^6.
\nonumber
\end{eqnarray}

\subsection{The ${\overline {\rm MS}}$  series up to  five loops}
\label{serms}

We report the perturbative series in the
$\overline{\rm MS}$ scheme.  The $\beta$ functions were computed in
Ref.~\cite{CP-04}, where they were explictly reported up to
three loops for generic U($M$)$\otimes$U($N$) models. Here we report
the five-loop series of the $\beta$ functions and of the RG functions
associated with the critical exponents
for the case relevant for QCD with two flavors,
i.e. the $\Phi^4$ theory (\ref{LUN}) with $N=2$.  

The $\beta$ functions are given by~\cite{footnoteseries} 
\begin{eqnarray}
&& \beta_u(u,v) = 
-u
+ 4 u^2+4 u v+\textstyle\frac{3}{2} v^2
 -\textstyle\frac{57}{8} u^3-11 u^2 v-\textstyle\frac{61}{8} uv^2 -
   3 v^3 
+ \textstyle\frac{93}{8} u^4 \zeta (3)+\textstyle\frac{389}{16} u^4+24 u^3 v \zeta (3)+
    \textstyle\frac{975}{16}  u^3 v 
\nonumber \\
 && +\textstyle\frac{99}{4} u^2 v^2 \zeta (3)+
     \textstyle\frac{9347}{128} u^2 v^2+
  18 u v^3  \zeta (3)+45 u v^3+6 v^4 \zeta (3)+
    \textstyle\frac{1197}{128} v^4
-\textstyle\frac{1885}{16} u^5 \zeta (5)
-\textstyle\frac{3119}{32} u^5 \zeta (3)
+\textstyle\frac{31}{120} \pi ^4  u^5
\nonumber \\
&& -\textstyle\frac{51759}{512} u^5-340 u^4 v \zeta (5)
-\textstyle\frac{1183}{4} u^4 v \zeta  (3)
+\textstyle\frac{161}{240} \pi ^4 u^4 v
-\textstyle\frac{10449 u^4}{32} v
-\textstyle\frac{3905}{8}  u^3 v^2 \zeta (5)
-\textstyle\frac{6849}{16} u^3 v^2 \zeta (3)
+\textstyle\frac{353}{480} \pi ^4  u^3 v^2
\nonumber \\
&&
-\textstyle\frac{391151}{768} u^3 v^2
-\textstyle\frac{895}{2} u^2 v^3 \zeta (5)-
377 u^2  v^3 \zeta (3)+\textstyle\frac{8}{15} \pi ^4 u^2 v^3
-\textstyle\frac{42919}{96} u^2 v^3
-\textstyle\frac{1875}{8} u v^4 \zeta (5)
-\textstyle\frac{3049}{16} u v^4 \zeta (3)
\nonumber \\
&&
+\textstyle\frac{31}{96} \pi ^4 u  v^4
-\textstyle\frac{12697}{64} u v^4
-50 v^5 \zeta (5)
-\textstyle\frac{325}{8} v^5 \zeta(3)
+\textstyle\frac{5}{48} \pi ^4 v^5
-\textstyle\frac{1097}{32} v^5
+ \textstyle\frac{646947}{512} u^6 \zeta (7)
+\textstyle\frac{333739}{256} u^6 \zeta (5)
\nonumber \\
&&
-\textstyle\frac{3}{64} u^6 \zeta (3)^2
+\textstyle\frac{333239}{512} u^6 \zeta (3)
-\textstyle\frac{1885}{4032} \pi ^6  u^6
-\textstyle\frac{6827}{2560} \pi ^4 u^6
+\textstyle\frac{121665}{256} u^6
+\textstyle\frac{146853}{32} u^5 v \zeta (7)
+\textstyle\frac{158151}{32} u^5 v \zeta (5)
\nonumber \\
&&
-\textstyle\frac{507}{32} u^5 v \zeta (3)^2
+\textstyle\frac{320791}{128} u^5 v \zeta (3)
-\textstyle\frac{6625}{4032} \pi ^6 u^5  v
-\textstyle\frac{37481}{3840} \pi ^4 u^5 v
+\textstyle\frac{494921}{256} u^5  v
+\textstyle\frac{4266675}{512} u^4 v^2 \zeta (7)
+\textstyle\frac{1144635}{128} u^4 v^2  \zeta (5)
\nonumber \\
&&
-\textstyle\frac{7017}{256} u^4 v^2 \zeta (3)^2
+\textstyle\frac{4795927}{1024} u^4 v^2 \zeta  (3)
-\textstyle\frac{9615}{3584} \pi ^6 u^4 v^2
-\textstyle\frac{504343}{30720} \pi ^4 u^4  v^2
+\textstyle\frac{7852881}{2048} u^4 v^2
+9702 u^3 v^3 \zeta (7)
+\textstyle\frac{79415}{8} u^3 v^3 \zeta (5)
\nonumber \\
&&
-\textstyle\frac{3}{4} u^3 v^3 \zeta  (3)^2
+\textstyle\frac{171801}{32} u^3 v^3 \zeta (3)
-\textstyle\frac{155}{56} \pi ^6 u^3  v^3
-\textstyle\frac{8173}{480} \pi ^4 u^3 v^3
+\textstyle\frac{6814435}{1536} u^3  v^3
+\textstyle\frac{3658095}{512} u^2 v^4 \zeta (7)
+\textstyle\frac{1773961}{256} u^2 v^4  \zeta (5)
\nonumber \\
&&
+\textstyle\frac{2777}{128} u^2 v^4 \zeta (3)^2
+\textstyle\frac{1942077}{512} u^2 v^4  \zeta (3)
-\textstyle\frac{94645}{48384} \pi ^6 u^2 v^4
-\textstyle\frac{182363}{15360} \pi ^4 u^2  v^4
+\textstyle\frac{36147287}{12288} u^2 v^4
+\textstyle\frac{189189}{64} u v^5 \zeta (7)
\nonumber \\
&&
+\textstyle\frac{44351}{16} u v^5 \zeta (5)
+\textstyle\frac{103}{8} u v^5 \zeta (3)^2
+\textstyle\frac{189841}{128} u v^5 \zeta (3)
-\textstyle\frac{2585}{3024} \pi ^6 u  v^5
-\textstyle\frac{1647}{320} \pi ^4 u v^5
+\textstyle\frac{2121643}{2048} u  v^5
\nonumber \\
&&
+\textstyle\frac{265041}{512} v^6 \zeta (7)
+\textstyle\frac{61459}{128} v^6 \zeta (5)
+\textstyle\frac{81}{64} v^6 \zeta (3)^2
+\textstyle\frac{246291}{1024} v^6 \zeta (3)
-\textstyle\frac{335}{2016} \pi ^6 v^6
-\textstyle\frac{3125}{3072} \pi ^4  v^6
+\textstyle\frac{2538035}{16384} v^6,
\end{eqnarray}
\begin{eqnarray}
&& \beta_v(u,v) = 
-v
+3u v
+2v^2
-\textstyle\frac{61}{8}u^2 v
-11u v^2
-\textstyle\frac{27}{8}v^3
+\textstyle\frac{1349}{64}u^3 v
+\textstyle\frac{1451}{32}u^2 v^2
+\textstyle\frac{575}{16}u v^3
+\textstyle\frac{347}{32}v^4
+\textstyle\frac{33}{2}u^3 v \zeta (3)
\nonumber \\
&&
+36u^2 v^2 \zeta (3)
+24u v^3 \zeta (3)
+\textstyle\frac{9}{2}v^4 \zeta (3)
-\textstyle\frac{49815}{512}u^4 v
+\textstyle\frac{29}{64}\pi ^4 u^4 v
-\textstyle\frac{27835}{96}u^3 v^2
+\textstyle\frac{163}{120}\pi ^4 u^3 v^2
-\textstyle\frac{272945}{768}u^2 v^3
+\textstyle\frac{22}{15}\pi ^4 u^2 v^3
\nonumber \\
&&
-\textstyle\frac{6635}{32}u v^4
+\textstyle\frac{53}{80}\pi ^4 u v^4
-\textstyle\frac{365}{8}v^5
+\textstyle\frac{1}{10}\pi ^4 v^5
-\textstyle\frac{3765}{32}u^4 v \zeta (3)
-\textstyle\frac{691}{2}u^3 v^2 \zeta (3)
-\textstyle\frac{6111}{16}u^2 v^3 \zeta (3)
-\textstyle\frac{1507}{8}u v^4 \zeta (3)
-\textstyle\frac{567}{16}v^5 \zeta (3)
\nonumber \\
&&
-\textstyle\frac{2625}{16}u^4 v \zeta (5)
-480u^3 v^2 \zeta (5)
-\textstyle\frac{2115}{4}u^2 v^3 \zeta (5)
-\textstyle\frac{1045}{4}u v^4 \zeta (5)
-\textstyle\frac{795}{16}v^5 \zeta (5)
+\textstyle\frac{445355}{1024}u^5 v
-\textstyle\frac{58367}{15360}\pi ^4 u^5 v
\nonumber \\
&&
-\textstyle\frac{12115}{16128}\pi ^6 u^5 v
+\textstyle\frac{209163}{128}u^4 v^2
-\textstyle\frac{109087}{7680}\pi ^4 u^4 v^2
-\textstyle\frac{7535}{2688}\pi ^6 u^4 v^2
+\textstyle\frac{16837765}{6144}u^3 v^3
-\textstyle\frac{81491}{3840}\pi ^4 u^3 v^3
-\textstyle\frac{8455}{2016}\pi ^6 u^3 v^3
\nonumber \\
&&
+\textstyle\frac{3808447}{1536}u^2 v^4
-\textstyle\frac{30289}{1920}\pi ^4 u^2 v^4
-\textstyle\frac{38005}{12096}\pi ^6 u^2 v^4
+\textstyle\frac{9331663}{8192}u v^5
-\textstyle\frac{1485}{256}\pi ^4 u v^5
-\textstyle\frac{28555}{24192}\pi ^6 u v^5
+\textstyle\frac{825245}{4096}v^6
-\textstyle\frac{1285}{1536}\pi ^4 v^6
\nonumber \\
&&
-\textstyle\frac{5}{28}\pi ^6 v^6
+\textstyle\frac{395479}{512}u^5 v \zeta (3)
+\textstyle\frac{734983}{256}u^4 v^2 \zeta (3)
+\textstyle\frac{282653}{64}u^3 v^3 \zeta (3)
+\textstyle\frac{56043}{16}u^2 v^4 \zeta (3)
+\textstyle\frac{736561}{512}u v^5 \zeta (3)
+\textstyle\frac{63485}{256}v^6 \zeta (3)
\nonumber \\
&&
+\textstyle\frac{2499}{128}u^5 v \zeta (3)^2
+\textstyle\frac{4305}{64}u^4 v^2 \zeta (3)^2
+\textstyle\frac{405}{4}u^3 v^3 \zeta (3)^2
+\textstyle\frac{2771}{32}u^2 v^4 \zeta (3)^2
+\textstyle\frac{2579}{64}u v^5 \zeta (3)^2
+\textstyle\frac{231}{32}v^6 \zeta (3)^2
+\textstyle\frac{105231}{64}u^5 v \zeta (5)
\nonumber \\
&&
+\textstyle\frac{390003}{64}u^4 v^2 \zeta (5)
+\textstyle\frac{295491}{32}u^3 v^3 \zeta (5)
+\textstyle\frac{457495}{64}u^2 v^4 \zeta (5)
+\textstyle\frac{5647}{2}u v^5 \zeta (5)
+\textstyle\frac{29035}{64}v^6 \zeta (5)
+\textstyle\frac{472311}{256}u^5 v \zeta (7)
\nonumber \\
&&
+\textstyle\frac{218295}{32}u^4 v^2 \zeta (7)
+\textstyle\frac{1325205}{128}u^3 v^3 \zeta (7)
+\textstyle\frac{1030617}{128}u^2 v^4 \zeta (7)
+\textstyle\frac{816291}{256}u v^5 \zeta (7)
+\textstyle\frac{16317}{32}v^6 \zeta (7).
\end{eqnarray}

The five-loop series of the RG functions
associated with the critical exponents are
\begin{eqnarray}
&& \eta_\phi(u,v) = 
+\textstyle\frac{5}{16}u^2
+\textstyle\frac{1}{2}u v
+\textstyle\frac{5}{16}v^2
-\textstyle\frac{5}{16}u^3
-\textstyle\frac{3}{4}u^2 v
-\textstyle\frac{93}{128}u v^2
-\textstyle\frac{1}{4}v^3 \nonumber \\
&& +\textstyle\frac{1125}{1024}u^4
+\textstyle\frac{225}{64}u^3 v
+\textstyle\frac{2835}{512}u^2 v^2
+\textstyle\frac{135}{32}u v^3
+\textstyle\frac{135}{128}v^4
-\textstyle\frac{485}{64}u^5
-\textstyle\frac{31}{1536}\pi^4 u^5
-\textstyle\frac{485}{16}u^4 v
-\textstyle\frac{31}{384}\pi^4 u^4 v
\nonumber \\
&& -\textstyle\frac{114469}{2048}u^3 v^2
-\textstyle\frac{119}{768}\pi ^4 u^3 v^2
-\textstyle\frac{28337}{512}u^2 v^3
-\textstyle\frac{31}{192}\pi ^4 u^2 v^3
-\textstyle\frac{57059}{2048}u v^4
-\textstyle\frac{1}{12}\pi ^4 u v^4
-\textstyle\frac{5743}{1024}v^5
-\textstyle\frac{31}{1920}\pi ^4 v^5
+\textstyle\frac{515}{512}u^5 \zeta(3)
\nonumber \\
&&
+\textstyle\frac{515}{128}u^4 v \zeta(3)
+\textstyle\frac{3593}{512}u^3 v^2 \zeta(3)
+\textstyle\frac{407}{64}u^2 v^3 \zeta(3)
+\textstyle\frac{1513}{512}u v^4 \zeta(3)
+\textstyle\frac{19}{32}v^5 \zeta(3),
\end{eqnarray}
\begin{eqnarray}
&& \eta_t(u,v) = 
-\textstyle\frac{5 u}{2}-2 v
+\textstyle\frac{15}{8}u^2
+3u v
+\textstyle\frac{15}{8}v^2
-\textstyle\frac{1195}{128}u^3
-\textstyle\frac{717}{32}u^2 v
-\textstyle\frac{2775}{128}u v^2
-\textstyle\frac{239}{32}v^3
+\textstyle\frac{9825}{256}u^4
+\textstyle\frac{31}{384}\pi ^4 u^4
+\textstyle\frac{1965}{16}u^3 v
\nonumber \\
&&
+\textstyle\frac{31}{120}\pi ^4 u^3 v
+\textstyle\frac{1431}{8}u^2 v^2
+\textstyle\frac{119}{320}\pi ^4 u^2 v^2
+\textstyle\frac{4029}{32}u v^3
+\textstyle\frac{31}{120}\pi ^4 u v^3
+\textstyle\frac{8103}{256}v^4
+\textstyle\frac{1}{15}\pi ^4 v^4
+\textstyle\frac{425}{64}u^4 \zeta (3)
+\textstyle\frac{85}{4}u^3 v \zeta (3)
\nonumber \\
&&
+\textstyle\frac{51}{2}u^2 v^2 \zeta (3)
+\textstyle\frac{107}{8}u v^3 \zeta (3)
+\textstyle\frac{209}{64}v^4 \zeta (3)
-\textstyle\frac{398085}{2048}u^5
-\textstyle\frac{2503}{6144}\pi ^4 u^5
-\textstyle\frac{9425}{96768}\pi ^6 u^5
-\textstyle\frac{398085}{512}u^4 v
-\textstyle\frac{2503}{1536}\pi ^4 u^4 v
\nonumber \\
&&
-\textstyle\frac{9425}{24192}\pi ^6 u^4 v
-\textstyle\frac{5888529}{4096}u^3 v^2
-\textstyle\frac{24209}{7680}\pi ^4 u^3 v^2
-\textstyle\frac{34885}{48384}\pi ^6 u^3 v^2
-\textstyle\frac{366513}{256}u^2 v^3
-\textstyle\frac{12767}{3840}\pi ^4 u^2 v^3
-\textstyle\frac{8705}{12096}\pi ^6 u^2 v^3
\nonumber \\
&&
-\textstyle\frac{11832549}{16384}u v^4
-\textstyle\frac{26999}{15360}\pi ^4 u v^4
-\textstyle\frac{17735}{48384}\pi ^6 u v^4
-\textstyle\frac{589857}{4096}v^5
-\textstyle\frac{691}{1920}\pi ^4 v^5
-\textstyle\frac{1795}{24192}\pi ^6 v^5
-\textstyle\frac{102955}{1024}u^5 \zeta (3)
-\textstyle\frac{102955}{256}u^4 v \zeta (3)
\nonumber \\
&&
-\textstyle\frac{364439}{512}u^3 v^2 \zeta (3)
-\textstyle\frac{85837}{128}u^2 v^3 \zeta (3)
-\textstyle\frac{334685}{1024}u v^4 \zeta (3)
-\textstyle\frac{16379}{256}v^5 \zeta (3)
+\textstyle\frac{4675}{256}u^5 \zeta (3)^2
+\textstyle\frac{4675}{64}u^4 v \zeta (3)^2
+\textstyle\frac{17687}{128}u^3 v^2 \zeta (3)^2
\nonumber \\
&&
+\textstyle\frac{4531}{32}u^2 v^3 \zeta (3)^2
+\textstyle\frac{9307}{128}u v^4 \zeta (3)^2
+\textstyle\frac{917}{64}v^5 \zeta (3)^2
+\textstyle\frac{85}{64}u^5 \zeta (5)
+\textstyle\frac{85}{16}u^4 v \zeta (5)
+\textstyle\frac{1601}{128}u^3 v^2 \zeta (5)
+\textstyle\frac{65}{4}u^2 v^3 \zeta (5)
\nonumber \\
&& +\textstyle\frac{461}{64}u v^4 \zeta (5)
-\textstyle\frac{1}{16}v^5 \zeta (5).
\end{eqnarray}

\end{widetext}


\begin{thebibliography}{99}
\bibitem{footnoteresum} The approximants we use depend on two parameters
  $\alpha$ and $b$;  we use here the notations of 
  Refs.~\cite{ZJ-book,CPV-00}.  In particular, 
  the RG trajectories reported in Figs.~\ref{mzmrgflow} and
  \ref{msrgflow} are obtained by using $\alpha=2$ and $b=5$
  for $\beta_u$ and $\alpha=1$ and $b=5$ for $\beta_v$ in both
  schemes.
\bibitem{footnoteBPV03} The first analysis of the MZM $\beta$
  functions, reported in Ref.~\cite{BPV-03}, was performed in the
  domain $-2\lesssim u \lesssim 4$, $0 < v \lesssim 4$, and found no
  evidence of a fixed point beside the O(8) FP that occurs in the
  model with $v_0 = 0$. Subsequent analyses \cite{BPV-05}, however,
  found an additional stable fixed point outside this region (point
  $S$ in Fig.~\ref{zeroesu22}), located at $u^* \approx -3.4$ and $v^* \approx
  5.3$.
\bibitem{footnote-err} The errors reflect the stability of the results
  of the different resummations with respect to the 
  parameters $b$ and $\alpha$ (see Ref.~\cite{footnoteresum}) 
  and the number of perturbative terms considered. They are obtained by 
  considering how the estimates change when $b$ and $\alpha$ 
  vary within a reasonably large region 
  around their optimal values (see Ref.~\cite{CPV-00} for a detailed
  description of the approach) 
  and when the resummation is applied to 
  the perturbative series with five (MZM scheme) or four ($\overline{\rm MS}$
  scheme) terms.
\bibitem{footnotecpv}
The estimate of the crossover exponent can be inferred from 
the results of Ref.~\cite{CPV-05}.
\bibitem{footnoteseries} We use a different normalization with respect to
Ref.~\cite{CP-04}: the quartic couplings we use differ by a
multiplicative factor of two from those of Ref.~\cite{CP-04}.
\end{thebibliography}
\end{document}